\documentclass[twocolumn]{aastex631}
\usepackage{amsmath}
\usepackage[T1]{fontenc}

\usepackage{xcolor}

\newcommand{\revise}[1]{{{#1}}}

\begin{document}

\title{Revisiting collisional dust growth in Class 0/I protostellar disks:\\Sweep-up can convert a few $10~{\rm M}_\oplus$ of dust into kg pebbles in 0.1~Myr}

\author[0000-0002-9408-2857]{Wenrui Xu}
\affiliation{Center for Computational Astrophysics, Flatiron Institute, New York, USA}

\author[0000-0001-5032-1396]{Philip J. Armitage}
\affiliation{Center for Computational Astrophysics, Flatiron Institute, New York, USA}
\affiliation{Department of Physics and Astronomy, Stony Brook University, Stony Brook, USA}

\begin{abstract}
Recent observations suggest that the first stages of planet formation likely take place in the Class 0/I phase of Young Stellar Object evolution, when the star and the disk are still embedded in an infalling envelope.
In this study we perform grain coagulation calculations to investigate the very first stage of planet formation, the collisional growth of dust grains, in Class 0/I disks.
We find that the slow increase in grain mass by high-velocity collision with much smaller grains (``sweep-up'') allows $\sim50~{\rm M}_\oplus$ of grains to grow well beyond the fragmentation barrier into $\sim$kg pebbles by the end of Class 0/I (0.1~Myr).
We analyze the linear growth and saturation of sweep-up to understand our results quantitatively, and test whether the sweep-up outcome is sensitive to disk parameters and details of the grain coagulation model.
The sweep-up pebble population could be important for planet formation, because they are less well-coupled to the gas (compared to the main population below the fragmentation barrier) and therefore more favorable to known mechanisms of dust clump formation (which initiate planetesimal formation).
It also contains enough mass to form all planet cores, based on observational estimates of the planet mass budget.
Our findings motivate future studies of grain growth and planetesimal formation in Class 0/I disks, including the subsequent evolution of this sweep-up population.
\end{abstract}


\section{Introduction}
Multiple lines of observational evidence now suggest that planet formation might yield Earth-mass and larger planetary cores during the Class 0/I phases of Young Stellar Object evolution.
Demographic studies of protoplanetary disks \citep{Manara22,Miotello22} show that the inferred dust mass in the Class II population must be converted into planets with order-unity efficiency if it is to form the known exoplanet population \citep{Najita14,Manara18,Dai2020,Lu2020,Mulders21}. 
The higher masses of Class 0/I disks \citep{Tychoniec20,Xu22}, conversely, imply nominal efficiencies of 2--30\% that appear more compatible with a lossy multi-step planet formation process.
More directly, substructure in ALMA-imaged disk systems \citep{Andrews20} is detected in disks with a wide range of ages, including some systems (HL~Tau, Elias~24) that are $\leq 1~{\rm Myr}$ old \citep{vanderMarel19}. If this substructure is interpreted as arising from planet-disk interaction \citep{Zhang18,Lodato19}, prompt growth to at least planetary core scales must be possible even at orbital radii of 10-100~au.
More recently, similar substructures are also detected in a small fraction of Class 0/I disks \citep{Sheehan2020}, although it is less clear whether they also originate from planet-disk interaction.

These observations call for theoretical studies of planet formation in Class 0/I disks.
Here, a natural first step would be revisiting known mechanisms of growing dust grains into planetary cores \citep[cf.][]{Lesur2022,Drazkowska2022} in the context of Class 0/I disks.
This is because previous studies of these mechanisms generally consider Class II disk environments, which could be very different from typical Class 0/I disk environments.
In particular, the infalling envelope around the protostar-disk system drives fast accretion (both from the envelope onto the disk and from the disk onto the star), and causes Class 0/I disks to have higher disk mass, temperature, and effective viscosity compared to their Class II counterparts (see discussion in Section \ref{sec:model:gas}).

This paper -- the first of a series of studies revisiting known mechanisms of growing dust grains into planetary cores in the context of Class 0/I disks -- focuses on the collisional growth of dust grains.
\nocite{XK21a} 
We perform a set of grain coagulation calculations based on established grain coagulation models \citep{B10,W12a} and a disk model that resembles typical Class 0/I disk properties \citep{XK21b}.
Our calculations suggest that ``sweep-up'', a mechanism for passing the fragmentation barrier by collisional growth that is found too inefficient in Class II disks \citep{W12a,W12b}, could instead form a sizeable population of large ($\sim$kg) pebbles in Class 0/I disks.
Using the results in this paper as initial conditions will also enable more self-consistent investigations of the subsequent stages of planet formation in future studies.

The rest of this paper is organized as follows. In Section \ref{sec:model} we introduce our Class 0/I disk model and dust coagulation model. In Section \ref{sec:results} we summarize and discuss the result of our grain coagulation calculation.
We discuss the linear growth and saturation of sweep-up and test whether the outcome is sensitive to disk properties and physical assumptions of the grain coagulation model.
In Section \ref{sec:discussion} we demonstrate why sweep-up is much more efficient in Class 0/I than in Class II, and argue that the sweep-up population could help subsequent planetesimal formation. In particular, it is even possible that the sweep-up polation alone accounts for most or all of planet core mass budget.
We summarize our results in Section \ref{sec:conclusion}.

\section{Model setup}\label{sec:model}
We perform dust coagulation calculations using the code \verb|dustpy| \citep{dustpy}, which solves gas and dust transport including viscous advection and diffusion as well as collisional growth and fragmentation of dust particles on a 1D (radial) disk profile.
Below we present the gas disk model and dust coagulation model in our calculations.

\subsection{Gas disk model for Class 0/I}\label{sec:model:gas}

We set up a simple 1D steady-state gas disk model with the goal of capturing the basic properties (density, temperature, level of turbulence) of a typical Class 0/I disk.
Because of theoretical and observational uncertainties regarding the exact time evolution of mass and angular-momentum accretion from the infalling envelope (or pseudodisk) onto the disk, we refrain from modeling the time evolution of the disk.

Following the physical insights from \citet{XK21b}, we assume that the angular-momentum transport in a typical Class 0/I disk is mainly provided by gravitational instability (GI) because there is no other efficient transport mechanism that can balance the accretion from envelope/pseudodisk infall.\footnote{In early studies of protostellar disk formation, it was generally assumed that angular-momentum transport in Class 0/I disks is due to GI \citep{AdamsLin1993}. Then the ``magnetic braking catastrophe'' in ideal MHD \citep{Allen2003} and early simulations adopting less realistic parametrizations of non-ideal MHD effects and/or relatively low resolution \citep[e.g.,][]{MellonLi2009, MachidaHosokawa2013} seem to suggest that magnetic effects (including magnetic braking and outflow launching) play an important role in regulating disk angular-momentum transport. On the other hand, high-resolution simulations with realistic models for ambipolar diffusion often find that ambipolar diffusion in the disk can be quite efficient and the disk can still be gravitationally unstable \citep[e.g.,][]{Masson2016, Zhao2018}, yet even in these studies GI often receives little attention in the discussion. More recently, \citet{XK21a,XK21b} demonstrated that the angular momentum transport inside the disk is dominated by the Reynolds and gravitational stresses associated with gravitationally excited spiral waves (while the envelope angular momentum transport, which sets the total angular momentum of the disk, is mainly magnetic). They also demonstrated that GI tends to spread radially in the disk, eventually making most part of the disk marginally unstable.}
We also assume that the disk is mainly heated by gravito-viscous accretion, instead of protostellar irradiation, even at $\sim 100$~au; this is because a marginally gravitationally unstable disk has a geometric thickness that peaks around the inner edge of the disk, which shields the outer disk from protostellar irradiation.
These assumptions have been shown to be consistent with simulation \citep{XK21b} and observations \citep{Zamponi2021,Xu22} of Class 0/I disks.

Given the physical constraints provided by these assumptions,
we can solve the radial profile of the disk at given stellar mass ($M_\star$) and accretion rate ($\dot M$, which we assume to be constant in radius).
First, steady-state accretion requires
\begin{equation}
    \dot M = 3\pi\Sigma\alpha c_s^2\Omega^{-1}.
    \label{eq:Mdot_alpha}
\end{equation}
Here $\Sigma$ is the surface density, $c_s$ is the sound speed, and $\Omega$ is the (Keplerian) rotation rate.\footnote{We assume that rotation is Keplerian in accordance with \texttt{dustpy}. This is slightly inconsistent with the fact that we have a massive disk though.}
$\alpha$ is the effective viscosity due to GI, which we model with a simple parametrization (following \citealt{Zhu2010}),
\begin{equation}
    \alpha = \exp(-Q^4).
\end{equation}
Here $Q = c_s\Omega/\pi G\Sigma$ is the Toomre $Q$ parameter.
Note that the resulting disk profile is insensitive to the exact choice of this parametrization as long as it captures the steep increase of $\alpha$ between the gravitationally stable regime ($\alpha\to 0$ for $Q\gtrsim 2$) and the highly unstable regime ($\alpha\sim 1$ for $Q\lesssim 1$).

A second constraint comes from thermal equilibrium
\begin{equation}
    \Lambda_{\rm heat} = \Lambda_{\rm cool},
    \label{eq:thermal_eq}
\end{equation}
where the heating $\Lambda_{\rm heat}$ is provided by gravito-viscous heating\footnote{This assumes that GI is exactly equivalent to a local viscosity. In reality, non-local energy and angular-momentum transport associated with self-gravitating large-scale spirals can change this result up to some $\mathcal O(1)$ factor \citep[][Appendix D]{XK21b}.}
\begin{equation}
    \Lambda_{\rm heat} = \frac{3\Omega^2}{4\pi}\dot M,
\end{equation}
and the cooling $\Lambda_{\rm cool}$ is provided by radiative cooling \citep{XK21b}
\begin{equation}
    \Lambda_{\rm cool} \sim \frac{8\tau}{1+0.875\tau^2} \sigma T^4.
\end{equation}
Here $\tau = \kappa\Sigma/2$ is the midplane optical depth, $\sigma$ is the Stefan-Boltzmann constant, and $T$ is the mean disk temperature (which is also approximately the midplane temperature). For simplicity, we do not distinguish between Rosseland and Planck mean opacities, and assume a very simple opacity profile
\begin{equation}
    \kappa(T) = \left\{
    \begin{array}{ll}
        \left(\frac{T}{100~{\rm K}}\right)^2 ~{{\rm cm}^2/{\rm g}} &  (T<100~{\rm K})\\
        1 ~{{\rm cm}^2/{\rm g}} & (T>100~{\rm K})
    \end{array}
    \right..
    \label{eq:kappa}
\end{equation}
Solving Eqs. \eqref{eq:Mdot_alpha} and \eqref{eq:thermal_eq} gives the radial profile of the disk ($\Sigma, T, \alpha$).

Our opacity table Eq. \eqref{eq:kappa} is a reasonable approximation of the actual dust opacity only up to $T_{\rm subl}\sim 1200~{\rm K}$; beyond $T_{\rm subl}$, most grains will sublime, and that leads to a decrease of opacity by a few orders of magnitude \citep{Pollack1994}.
Since the disk is optically thick when midplane temperature reaches $T_{\rm subl}$, grain sublimation around the midplane makes cooling more efficient and pins the disk at $T_{\rm subl}$. Therefore, when Eqs. \eqref{eq:Mdot_alpha} and \eqref{eq:thermal_eq} yield $T>T_{\rm subl}$, we set $T$ to $T_{\rm subl}$ and solve Eq. \eqref{eq:Mdot_alpha} for $\alpha$ and $\Sigma$.

\begin{figure}
    \centering
    \includegraphics[scale=0.66]{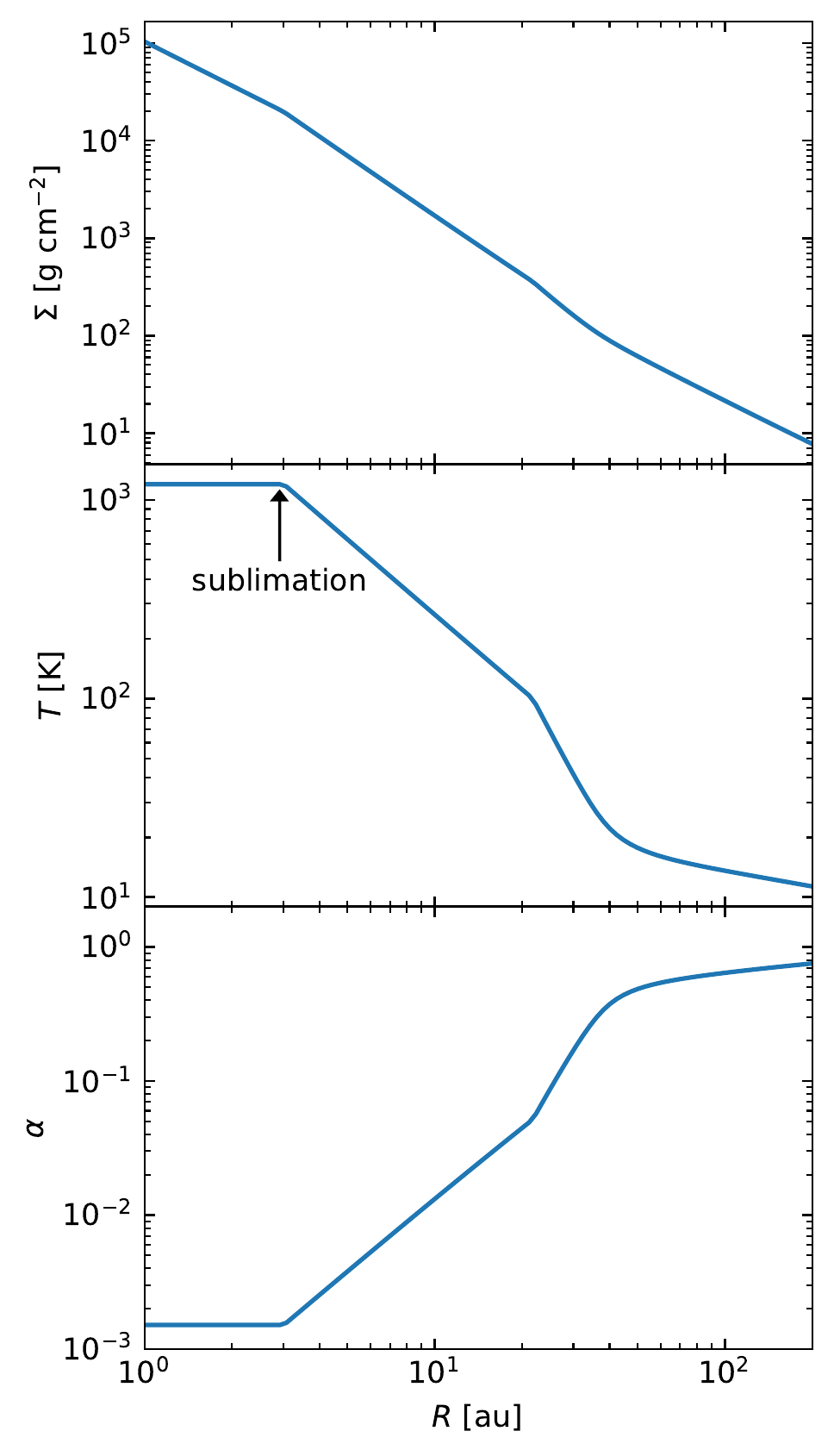}
    \caption{The Class 0/I protostellar disk model used for our calculation. This model corresponds to a gravitationally self-regulated steady-state disk with $M_\star=0.5~{\rm M}_\odot$, $R_{\rm d}=200~{\rm au}$, and $\dot M=10^{-5}~{\rm M}_\odot/{\rm yr}$.}
    \label{fig:disk}
\end{figure}

For our fiducial model, we choose a stellar mass of $M_\star = 0.5{\rm M}_\odot$, and an accretion rate of $\dot M = 10^{-5}~{\rm M}_\odot/{\rm yr}$ based on the typical duration of the main accretion phase ($\sim 10^5~{\rm yr}$).
Fig. \ref{fig:disk} shows the radial profile of our fiducial gas disk's surface density $\Sigma$, temperature $T$, and effective viscosity $\alpha$.
We also explore the effect of choosing different disk parameters in Section \ref{sec:results:sensitivity_disk}.

\subsection{Dust coagulation model}\label{sec:model:dust}

Our dust coagulation model is based on the default model in \texttt{dustpy}, with a few modifications to capture the mass transfer during collision more realistically and to model grain sublimation at $T_{\rm subl}$. Below we review the default model in \texttt{dustpy} and introduce our modifications. We also discuss important uncertainties in modeling dust coagulation in Appendix \ref{a:uncertainties}.

The default grain coagulation model in \texttt{dusty} is mainly based on \citet{B10}.
For two dust grains, labeled by projectile (lower mass, $m_{\rm p}$) and target (higher mass, $m_{\rm t}$), colliding at relative velocity $v_{\rm rel}$, the model includes three different outcomes of grain-grain collision.
When relative velocity is low ($v_{\rm rel} < 1~{\rm m/s}$; cf. Appendix \ref{a:uncertainties:frag}), the two grains coagulate.
When relative velocity is high and the two grains have similar masses ($v_{\rm rel} > 1~{\rm m/s}$, $m_{\rm t}/m_{\rm p}<10$), a full fragmentation occurs, and the mass in both grains is turned into a power-law fragment distribution with maximum mass $m_{\rm t}$.
When relative velocity is high and the two grains have very different masses ($v_{\rm rel} > 1~{\rm m/s}$, $m_{\rm t}/m_{\rm p}>10$), an erosion event occurs, where the projectile fragments and excavates $m_{\rm er}$ from the target; we are then left with a new target mass of $m_{\rm t}-m_{\rm er}$, and the remaining mass ($m_{\rm p}+m_{\rm er}$) is turned into a power-law fragment distribution with maximum mass $m_{\rm p}$. By default \texttt{dustpy} assumes an erosion efficiency of unity, $m_{\rm er} = m_{\rm p}$.

Under this model, the evolution of the grain size is fairly simple. If we ignore the radial evolution, grain growth can only proceed up to the fragmentation barrier (where typical grain velocity is $\sim 1~{\rm m/s}$). Further collisional growth is impossible because large grains above the fragmentation barrier are constantly fragmented or eroded by collision with other grains.

\subsubsection{Modeling mass transfer}

Collisional growth beyond the fragmentation barrier becomes possible when one includes a more realistic model of mass transfer during collision. \citet{W12a} developed a velocity-dependent mass transfer model based on lab experiments, and found that there is a parameter regime where a high-velocity collision still allows the target to gain a small amount of mass by having some of the fragmented projectile sticking onto the target. Therefore, if one introduces a small amount of large grains into a population of small grains, the large grains could slowly grow by ``sweeping up'' the smaller grains.

Here we make a simple adjustment to our grain coagulation model based on \citet{W12a} to model this sweep-up mechanism.
Our modification only concerns the erosion regime in \texttt{dustpy}'s fiducial model.
In this regime, instead of having the target mass changed to $m_{\rm t}-m_{\rm er}$ with $m_{\rm er}=m_{\rm p}$, we now set the post-collision target mass to $m_{\rm t}-m_{\rm er}+m_{\rm mt}$, where the erosion mass ($m_{\rm er}$) is given by (\citealt{W12a} Eqs. 17-18)
\begin{equation}
    \frac{m_{\rm er}}{m_{\rm p}} = \revise{9.3\times 10^{-6}}\left(\frac{m_{\rm p}}{3.5\times 10^{-12}~{\rm g}}\right)^{0.15}\frac{v_{\rm rel}}{1~{\rm cm/s}} - 0.4,\label{eq:m_er}
\end{equation}
and the mass transfer ($m_{\rm mt}$) is given by (\citealt{W12a} Eq. 14)
\begin{equation}
    \frac{m_{\rm mt}}{m_{\rm p}} = -6.8\times 10^{-3} + 3.6\times 10^{-5}\frac{v_{\rm rel}}{1~{\rm cm/s}}.\label{eq:m_mt}
\end{equation}
We cap both $m_{\rm er}$ and $m_{\rm mt}$ to $[0, m_{\rm p}]$. The resulting fragments, which have a total mass of $m_{\rm p}-m_{\rm mt}+m_{\rm er}$, are distributed into a power-law distribution with maximum fragment $m_{\rm p}$, following the default \texttt{dustpy} model in the erosion regime.


\revise{One caveat is that these parametrizations of erosion and mass transfer might not be sufficiently realistic. In particular, our parametrization of erosion efficiency could have underestimated the contribution from small grains, while the possible difference in the compactness of projectile and target may enhance mass transfer and suppress erosion. We present a more detailed discussion of these uncertainties in Appendix \ref{a:uncertainties:mt}.}

\subsubsection{Modeling sublimation}
In the innermost part of the disk, the midplane temperature reaches $T_{\rm subl}$ and nearly all dust grains should sublime.
In order to model grain sublimation at $T_{\rm subl}$, in the beginning of each timestep we convert all dust mass in the region where $T\geq T_{\rm subl}$ (which in our fiducial model corresponds to $r\lesssim 3~{\rm au}$) into small grains that follow a MRN size distribution (${\rm d}n/{\rm d}a \propto a^{-3.5}$; \citealt{MRN77}) with maximum grain size $a_{\rm max}=1~\mu$m. Here we use these small (Stokes number ${\rm St}\ll 1$ and well below fragmentation barrier) grains as a proxy of grain material in gas phase; it is important to keep their mass in the calculation because gas-phased grain material can still diffuse outwards and condensate into solid again.

In this study we do not consider the partial sublimation of grain materials at lower temperatures (e.g. snowline) and the resulting changes in grain mass and grain properties. We leave the effect of snowline, which might be important for planet formation \citep{Drazkowska2017}, to future studies.

\subsection{Other simulation setup}
We use a log-uniform radial grid that spans $[2~{\rm au}, R_{\rm d}]$ with default disk size $R_{\rm d}=200~{\rm au}$ and a resolution of 6 cells per factor of 2,
and a log-uniform grain mass grid that spans $[10^{-13}, 10^{13}]~{\rm g}$ with a resolution of 7 cells per order of magnitude. (The large maximum grain mass is mostly precautionary; in our simulations we never achieve such high grain mass.)
We also experiment with smaller disk sizes in Section \ref{sec:results:sensitivity_disk}.

The dust initial condition is given by a constant dust-to-gas ratio of 0.01 with a MRN size distribution (${\rm d}n/{\rm d}a \propto a^{-3.5}$; \citealt{MRN77}) with maximum grain size $a_{\rm max}=1~\mu$m.
The inner boundary condition for dust evolution is a constant gradient boundary condition, following \texttt{dustpy}'s default setup.
The outer boundary condition for dust evolution is given by a fixed dust profile in the outermost (ghost) cell with a dust-to-gas ratio of 0.01 and MRN size distribution with $a_{\rm max}=1~\mu$m (same as the initial condition).\footnote{In reality, $1~\mu$m is not necessarily a good estimate of $a_{\rm max}$ in the envelope/pseudodisk, but since the grain growth up to fragmentation barrier happens quickly compared to the timescale of evolution, adopting a somewhat arbitrary envelope grain size would barely affect the results in the disk.}
This treatment captures the grain mass flux into the disk from the surrounding envelope/pseudodisk. Note that since $a_{\rm max}=1~\mu$m still corresponds to ${\rm St}\ll 1$, the ratio between dust and gas mass flux from the envelope onto the disk is the same as the dust-to-gas ratio in the envelope (0.01).

For our fiducial calculation we integrate the system for $10^{5}~{\rm yr}$; at the end of the calculation, the result could be interpreted as a rough estimate of the dust properties when Class 0/I (or the main accretion phase) ends, or the initial conditions of Class II (protoplanteary) disk evolution, with the caveat that we do not include the decrease in disk mass expected during the transition into Class II.

\section{Results}\label{sec:results}
\begin{figure*}
    \centering
    \includegraphics[scale=0.66]{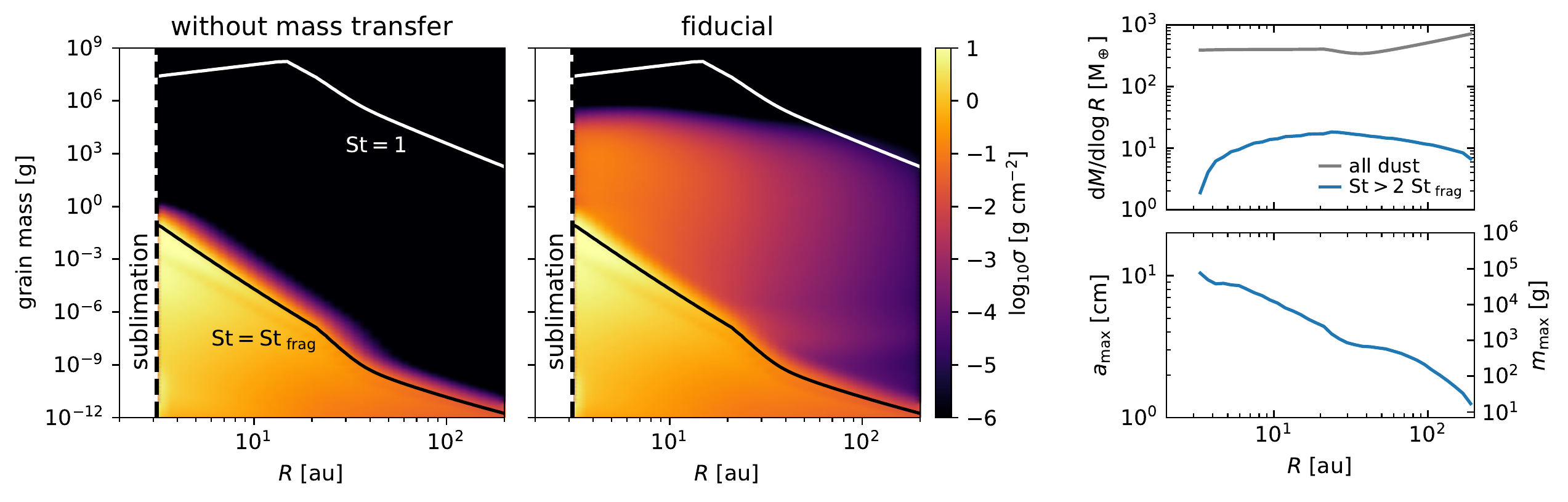}
    \caption{Summary of our fiducial model at $t=10^5~{\rm yr}$ (which roughly corresponds to end of Class 0/I). Left panels: dust surface density $\sigma$ (with $\sigma\equiv {\rm d}\Sigma/{\rm d}\log m$) as a function of radius $R$ and grain mass $m$. A calculation without mass transfer is also shown for reference. Top right panel: mass in all dust grains and the sweep-up population. Here we define the sweep-up population as large grains well above the fragmentation barrier, with ${\rm St}>2{\rm St}_{\rm frag}$. The total mass in the sweep-up population is $\revise{52}~{\rm M}_\oplus$. Bottom right panel: Maximum grain size $a_{\rm max}$ and grain mass $m_{\rm max}$. These are defined as the 95th percentile of the sweep-up population.}
    \label{fig:overview}
\end{figure*}

The result of our fiducial model is summarized in Fig. \ref{fig:overview}. Compared to a model without mass transfer, we see a new population of large grains (more precisely, pebbles) emerging above the fragmentation barrier due to sweep-up.
Here we define the fragmentation barrier ${\rm St}_{\rm frag}$ as the Stokes number (which is a proxy of grain size) where sticking probability equals fragmentation probability for same-sized particles.
Using ${\rm St}_{\rm frag}$, we separate the dust mass into two populations, a main population with ${\rm St}<2{\rm St}_{\rm frag}$, and a sweep-up population with ${\rm St}>2{\rm St}_{\rm frag}$.
We include the prefactor 2 because a small tail of the main population is expected to extend slightly beyond the fragmentation barrier (cf. Fig. \ref{fig:overview} left panel).

In our fiducial run, the sweep-up population has a total mass of $\revise{52}~{\rm M}_\oplus$ and is distributed mainly at 4--200 au. These large grains attain a maximum size of $1$--$10~{\rm cm}$ and mass $1~{\rm g}$--$100~{\rm kg}$, depending on the radius.

In the remaining of this section, we try to quantitatively understand the origin of this sweep-up population of large pebbles, including what sets its grain mass and surface density.
This is done in three steps.
We first analyze the sweep-up growth rate, identifying parameter regimes where growth is or is not possible and calculating the scaling laws of sweep-up growth.
We then discuss what limits sweep-up growth and sets the mass budget of large grains in the absence of radial dust evolution (advection, drift, and radial diffusion).
Finally, we discuss the effect of radial dust evolution.
We also test whether the results are sensitive to changes in disk parameter and dust coagulation model.

\subsection{Sweep-up growth rate}\label{sec:results:growth}

\begin{figure}
    \centering
    \includegraphics[scale=0.66]{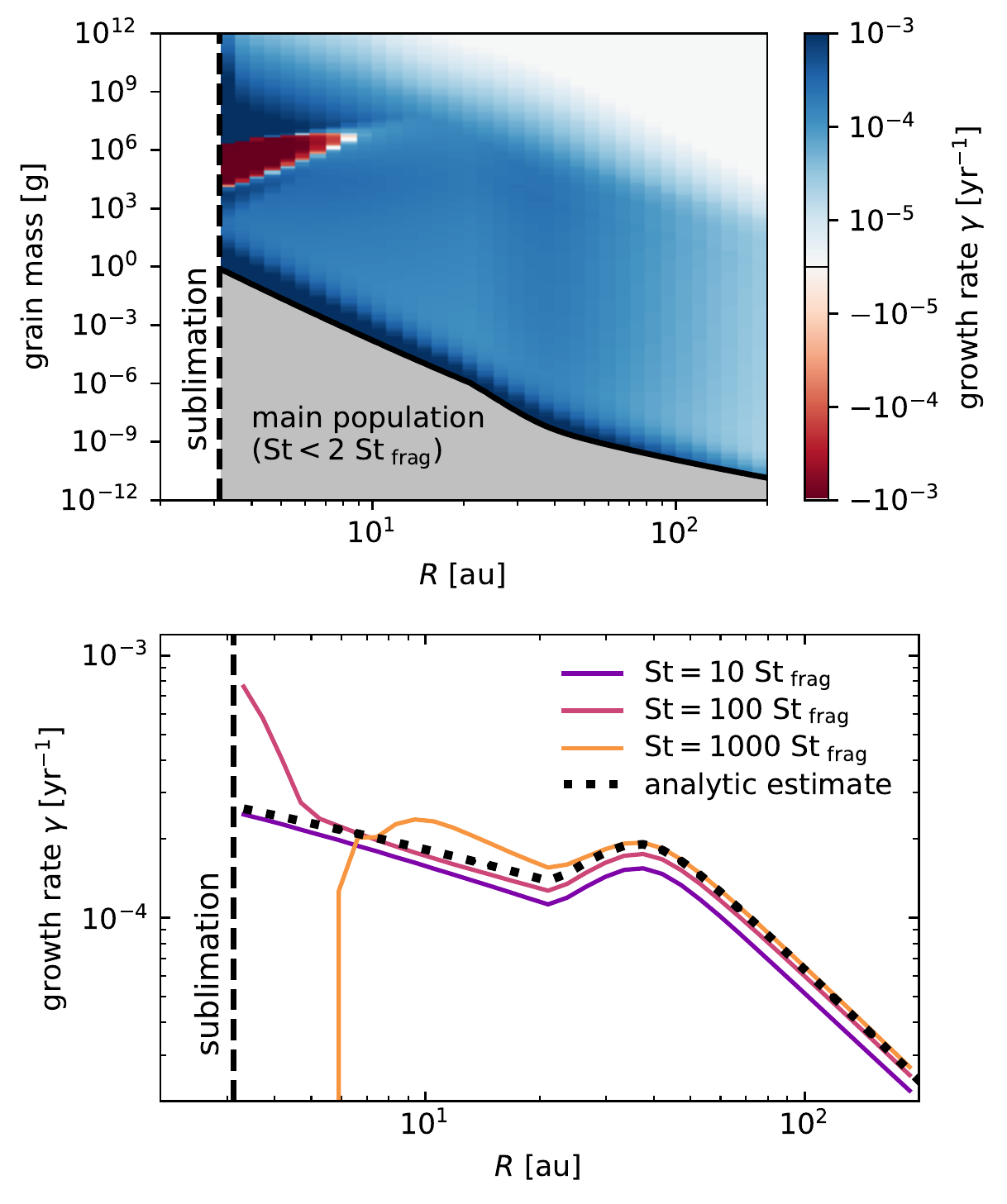}
    \caption{Top panel: sweep-up growth rate $\gamma$, computed following Eq. \eqref{eq:gamma}. Mass transfer results in net growth in most cases, except for an erosion zone at small radii. Bottom panel: growth rate at different ${\rm St}$ (a proxy of grain size), compared with the simple analytic estimate in Eq. \eqref{eq:gamma_simple}.
    At a given radius far from the erosion zone, the growth rate is insensitive to grain size.}
    \label{fig:growth_rate}
\end{figure}

In Fig. \ref{fig:growth_rate}. we plot the growth rate by sweep-up measured from our fiducial simulation. The growth rate can be defined by
\begin{equation}
    \gamma = \int \frac{m_{\rm mt}-m_{\rm er}}{m_{\rm t}} \langle\sigma v_{\rm rel}\rangle \frac{{\rm d}n}{{\rm d}m_{\rm p}} {\rm d} m_{\rm p}.\label{eq:gamma}
\end{equation}
Here $\frac{m_{\rm mt}-m_{\rm er}}{m_{\rm t}}$ is the fractional net mass change per collision due to mass transfer (we take it to be zero when $m_{\rm t}<10m_{\rm p}$), $\langle\sigma v_{\rm rel}\rangle$ is the collision rate, and $n$ is the number density of the main population (${\rm St}<2{\rm St}_{\rm frag}$).

Mass transfer gives positive growth rate in most cases, except in an ``erosion zone'' at small radii. The existence of the erosion zone can be understood from the erosion and mass transfer efficiency in Eqs. \eqref{eq:m_er} and \eqref{eq:m_mt}. In the limit of high relative velocity, the net change in mass is
\begin{equation}
    m_{\rm mt}-m_{\rm er} \approx  \epsilon_{\rm net}\frac{v_{\rm rel}}{1~{\rm cm/s}}m_{\rm p},
\end{equation}
where
\begin{equation}
    \epsilon_{\rm net} \equiv 3.6\times 10^{-5} - \revise{9.3\times 10^{-6}}\left(\frac{m_{\rm p}}{3.5\times 10^{-12}{\rm g}}\right)^{0.15}.\label{eq:eps_net}
\end{equation}
This change is positive when $m_{\rm p}< 10^{-8}{\rm g}$. Therefore, at larger radii where most grains in the main population are small, mass transfer always leads to net growth. Meanwhile, at smaller radii where most grains are \revise{$> 3\times 10^{-8}{\rm g}$}, mass transfer can result in net erosion. In our disk, the median grain size of the main population falls below \revise{$3\times 10^{-8}{\rm g}$} at $\sim20~{\rm au}$. In reality the radial boundary of the erosion zone is slightly smaller than that (at $\sim 10~{\rm au}$) mainly because the other terms in Eqs. \eqref{eq:m_er} and \eqref{eq:m_mt} has a net effect of increasing $m_{\rm mt}-m_{\rm er}$.

Another interesting feature in Fig. \ref{fig:growth_rate} is that at a given radius that is far from the erosion zone, the growth rate generally remains approximately constant for many orders of magnitude before decreasing at large grain size.
This result can be explained by estimating the growth rate in the limit of small $m_{\rm p}$. In this limit, we have
\begin{equation}
    \epsilon_{\rm net} \sim 3.6\times 10^{-5},~~~ \sigma\sim \pi a_{\rm t}^2,~~~ v_{\rm rel} \sim v_{\rm t}.
\end{equation}
Here $a_{\rm t}$ is the target size and $v_{\rm t}$ is the target velocity with respect to gas. Under these approximations, Eq. \ref{eq:gamma} can be simplified into
\begin{equation}
    \gamma \sim \pi \epsilon_{\rm net}(1~{\rm cm/s})^{-1} \rho_{\rm d} a_{\rm t}^2 v_{\rm t}^2 m_{\rm t}^{-1}.
\end{equation}
Here $\rho_{\rm d}$ is the midplane dust mass density.
Now consider $v_{\rm t}$. In our model $v_{\rm t}$ is dominated by turbulent velocity, which is (\citealt{OrmelCuzzi2007}, Eq. 28 and below, Eq. 29; also see \citealt{B12} Eq. 11)
\begin{equation}
    v_{\rm t}^2 \sim \min\left\{\frac 92 \alpha c_s^2 {\rm St}, \frac 32 \alpha c_s^2 \left(1+\frac{1}{1+{\rm St}}\right)\right\}.\label{eq:vt}
\end{equation}
Meanwhile, under typical disk conditions St is simply (\citealt{B12} Eq. 2)
\begin{equation}
    {\rm St} = \frac{a\rho_{\rm s}}{\Sigma_{\rm g}}\frac{\pi}{2}.
\end{equation}
Here $\rho_{\rm s}$ is the density of grain material and $\Sigma_{\rm g}$ is the gas surface density.
When St is not too large so Eq. \eqref{eq:vt} takes the first term, the dependence of $\gamma$ on target properties all cancel out:
\begin{align}
    \gamma \sim& \frac{27}{16}\sqrt{\frac{\pi}{2}} \epsilon_{\rm net} (1~{\rm cm/s})^{-1} \frac{\rho_{\rm d}}{\rho_{\rm g}}\Omega \alpha c_s \nonumber\\
    \sim& 3\times 10^{-5} \left(\frac{\rho_{\rm d}/\rho_{\rm g}}{0.01}\right) \left(\frac{M_\star}{0.5~{\rm M}_\odot}\right)^{1/2} \left(\frac{R}{20~{\rm au}}\right)^{-3/2}\nonumber \\
    & \left(\frac{\alpha}{0.01}\right) \left(\frac{T}{100~{\rm K}}\right)^{1/2} {\rm yr}^{-1}.\label{eq:gamma_simple}
\end{align}
Here $\rho_{\rm d}/\rho_{\rm g}$ is the midplane dust-to-gas ratio.
This gives a growth rate that is approximately constant in target grain mass before St becomes large enough to make Eq. \eqref{eq:vt} take the second term.

\subsection{Sweep-up outcome in the absence of radial evolution}\label{sec:results:norad}

\begin{figure}
    \centering
    \includegraphics[scale=0.66]{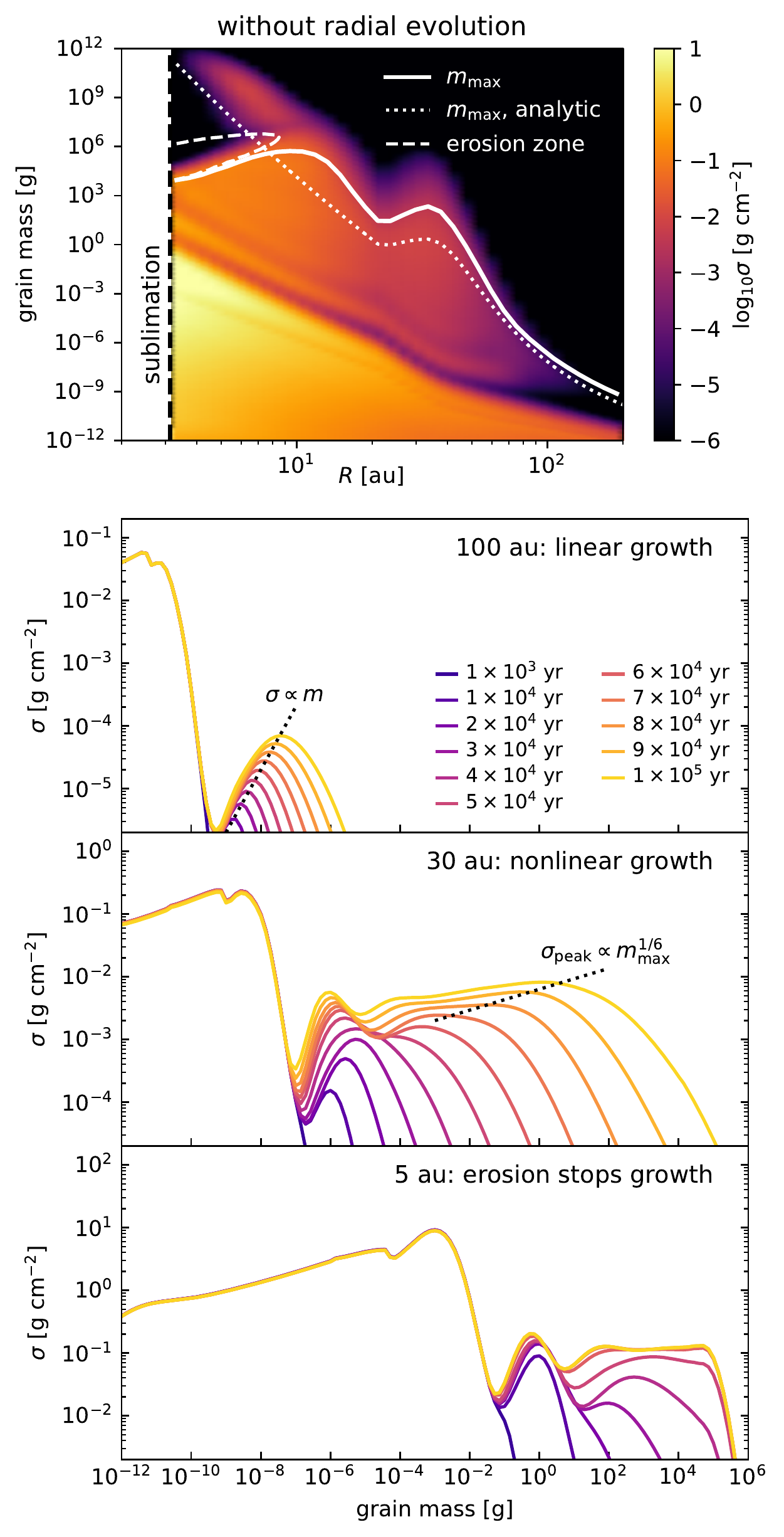}
    \caption{Top panel: dust distribution of a run without radial dust evolution at $t=10^5~{\rm yr}$. We also show the maximum grain mass $m_{\rm max}$, its analytic estimate assuming linear growth [Eq. \eqref{eq:m_max_estimate}], and the boundary of the erosion zone (Fig. \ref{fig:growth_rate}) for reference. $m_{\rm max}$ is well approximated by the analytic growth limit at large radii, and the erosion zone boundary at small radii.
    Bottom panels: time evolution of grain size distribution at radii representative of different regimes of growth.}
    \label{fig:no_rad_analysis}
\end{figure}

Now we consider the outcome of sweep-up. To simplify the problem, in this subsection we consider a case without dust radial evolution (advection, drift, and diffusion). This makes the evolution at a given radius completely local. The effect of dust radial evolution will be discussed in the next subsection.

The result of sweep-up growth without radial evolution is summarized in Fig. \ref{fig:no_rad_analysis}. Depending on the radius, there are several different regimes of evolution (bottom panels of Fig. \ref{fig:no_rad_analysis}), and we discuss each of them below.

\textbf{Large radii -- linear growth:}
At large radii (100~au panel of Fig. \ref{fig:no_rad_analysis}), the sweep-up population attains a self-similar profile with $\sigma\propto m$. This is because each large grain increases its mass at a similar rate ($\gamma$), resulting in a profile with approximately constant number density per $\log m$. The maximum grain size can be roughly estimated using the growth rate,
\begin{equation}
    m_{\rm max}(t)\sim m_{\rm init} e^{\gamma t}.\label{eq:m_max_estimate}
\end{equation}
Here $\gamma$ is the sweep-up growth rate [Eqs. \eqref{eq:gamma} and \eqref{eq:gamma_simple}], and $m_{\rm init}$ is the initial grain mass for sweep-up growth, which we empirically take as the mass at ${\rm St}=2~{\rm St}_{\rm frag}$. As shown in the top panel of Fig. \ref{fig:no_rad_analysis}, this estimate is in good agreement with the actual $m_{\rm max}$.

The amplitude of this self-similar profile depends on the intercept $\sigma/m$, which is set by the tail of the main population that extends beyond the fragmentation barrier; more precisely, by the tail's amplitude when sweep-up growth exceeds fragmentation with similar-sized grains. This tail exists because the relative velocity between two grains is not fixed but follows a Maxwellian distribution, and there is always a finite possibility of coagulation by low-velocity collision.
Empirically, it is usually several orders of magnitude below the main population when sweep-up growth begins, and this might be sensitive to the details of the grain coagulation model \citep{W12b}.

\textbf{Intermediate radii -- nonlienar growth:}
The linear growth discussed above does not proceed indefinitely. Eventually, $\sigma$ at the peak of the sweep-up population becomes large enough to make fragmentation by collision with similar-sized grains non-negligible.

Unlike the linear growth regime, the surface density of the sweep-up population is no longer sensitive to how much seeding is provided by the tail of the main population.
Instead, it is set by a balance between growth and fragmentation. Equating the approximately constant growth rate with the rate of fragmentation (of a given large grain) with similar-sized grains, which is $\propto (\sigma/m) a^2 v_{\rm rel}\propto \sigma m^{-1/6}$ (cf. Eq.~\ref{eq:vt}), we expect the peak $\sigma$ of the sweep-up population to scale with maximum grain size as
\begin{equation}
    \sigma_{\rm peak} = \sigma(m_{\rm max}) \propto m_{\rm max}^{1/6}.
\end{equation}
This is indeed broadly consistent with the time evolution shown in the 30~au panel of Fig. \ref{fig:no_rad_analysis}.

We also find that the growth of $m_{\rm max}$ is still reasonably approximated by Eq. \eqref{eq:m_max_estimate}. In other words, sweep-up growth and fragmentation can be considered as two independent effects: the first increases grain size while conserving number density, and the latter reduces the number density at the peak (and turns the corresponding mass into smaller grains) but does not directly affect maximum grain size.

\textbf{Small radii -- growth capped by erosion:} The last regime is similar to the previous nonlinear growth regime, except that the growth of $m_{\rm max}$ is now set by the erosion zone. We also comment that the sweep-up population tends to develop a tail that penetrates the erosion zone (similar to how the main population penetrates the fragmentation barrier), and could eventually lead to a small population of very large grains beyond the erosion zone. Such population is visible at $R\lesssim 10~{\rm au},~m\gtrsim10^6~{\rm g}$ in the top panel of Fig. \ref{fig:no_rad_analysis}).

\subsection{The effect of radial evolution}\label{sec:results:rad}

\begin{figure}
    \centering
    \includegraphics[scale=.66]{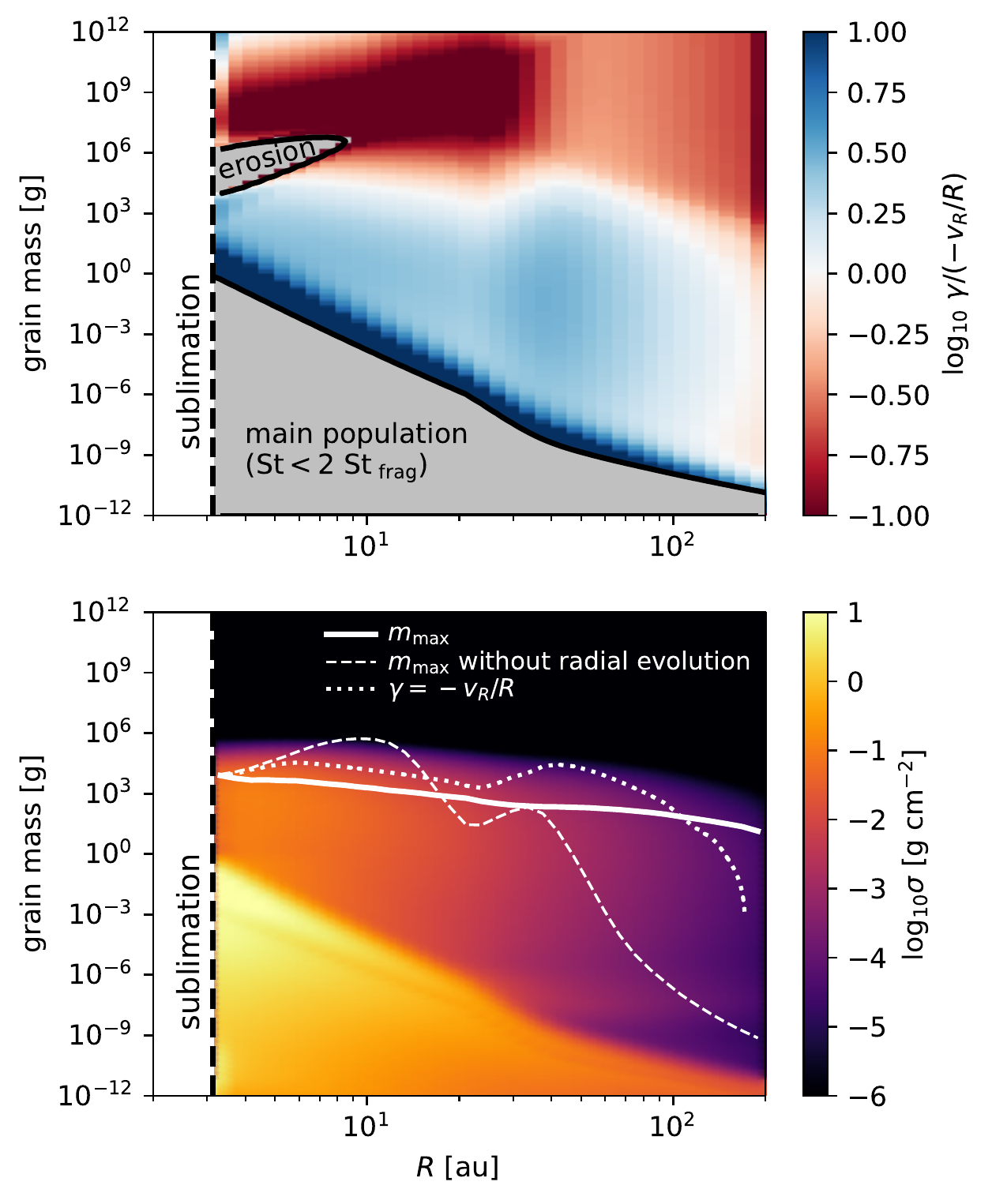}
    \caption{Top panel: ratio between sweep-up growth rate $\gamma$ and the rate of radial motion $-v_R/R$. Here $v_R$ includes both advection and drift. Bottom panel: dust distribution of our fiducial run at $t=10^5~{\rm yr}$. We also plot the maximum grain mass $m_{\rm max}$, $m_{\rm max}$ from the run without radial evolution in Fig. \ref{fig:no_rad_analysis}, and the grain mass where growth balances radial motion. At small radii where growth is fast, grain growth is capped by radial motion. At large radii where growth is slow, grain growth is seeded by the outward turbulent diffusion from smaller radii, which produces larger grains and more mass in the sweep-up population compared to the run without radial evolution.}
    \label{fig:rad_analysis}
\end{figure}

Finally, we are ready to understand the results of our fiducial run by considering the effect of radial evolution on top of sweep-up growth (Fig. \ref{fig:rad_analysis}). The radial evolution of dust grains have two main effects: First, it removes grains via inward advection and drift; second, it diffuses grains outward (because at given grain size, dust-to-gas ratio usually decreases outwards) via turbulent diffusion. These two effects shape the evolution differently at small and large radii.

\textbf{Small radii -- growth capped by radial motion:} As grain size increases, the rate of radial motion increases (due to an increase in drift velocity) and the growth rate decreases; therefore, there is a critical grain mass where radial motion can cap growth ($\gamma=-v_R/R$; cf. top panel of Fig. \ref{fig:rad_analysis}).
At small radii, dust growth proceeds quickly and this critical mass is reached within the lifetime of the system. Therefore we have grain growth capped around $\gamma=-v_R/R$, and the evolution is otherwise similar to the nonlinear growth regime discussed previously (cf. Fig. \ref{fig:rad_analysis} bottom panel, $R\lesssim 15~{\rm au}$). This can reduce $m_{\rm max}$ compared to the case without radial evolution.

\textbf{Large radii -- growth seeded by outward diffusion:} In the absence of radial dust evolution, the growth at large radii is mainly limited by the seeding of large grains by the tail of the main population, which is low in both grain size and number density.
The inclusion of radial evolution causes outward turbulent diffusion of large grains, which now allows the sweep-up growth to begin at higher grain mass and number density in the outer disk.
This produces much higher $\sigma$ and $m_{\rm max}$ in the sweep-up population at large radii compared to the case without radial evolution (Fig. \ref{fig:rad_analysis} bottom panel, $R\gtrsim 50~{\rm au}$).

In summary, at small radii radial evolution suppresses sweep-up via advection and drift, and at large radii radial evolution promotes sweep-up by seeding growth with outward diffusion. This leads to the smooth profile of large grains in our fiducial simulation (cf. Fig. \ref{fig:overview} top right panel).

\subsection{Sensitivity to disk parameters}\label{sec:results:sensitivity_disk}
\begin{figure}
    \centering
    \includegraphics[scale=0.66]{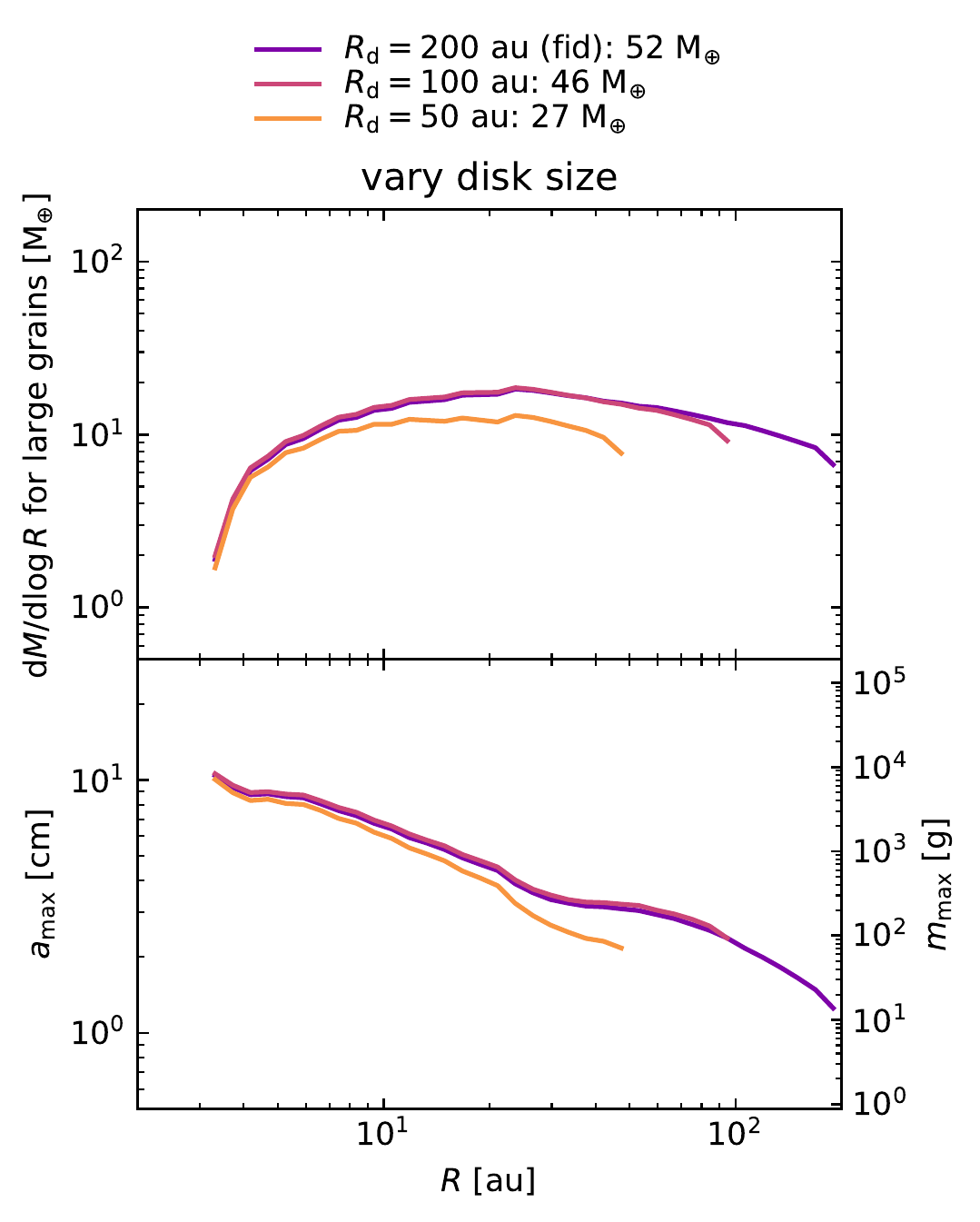}
    \caption{Comparison of the sweep-up population's mass and maximum grain size/mass for different disk size $R_{\rm d}$. We also mark the total mass in the sweep-up population in the legend.
    Choosing a different disk size barely affects the radial profiles.}
    \label{fig:vary_Rd}
\end{figure}
\begin{figure}
    \centering
    \includegraphics[scale=0.66]{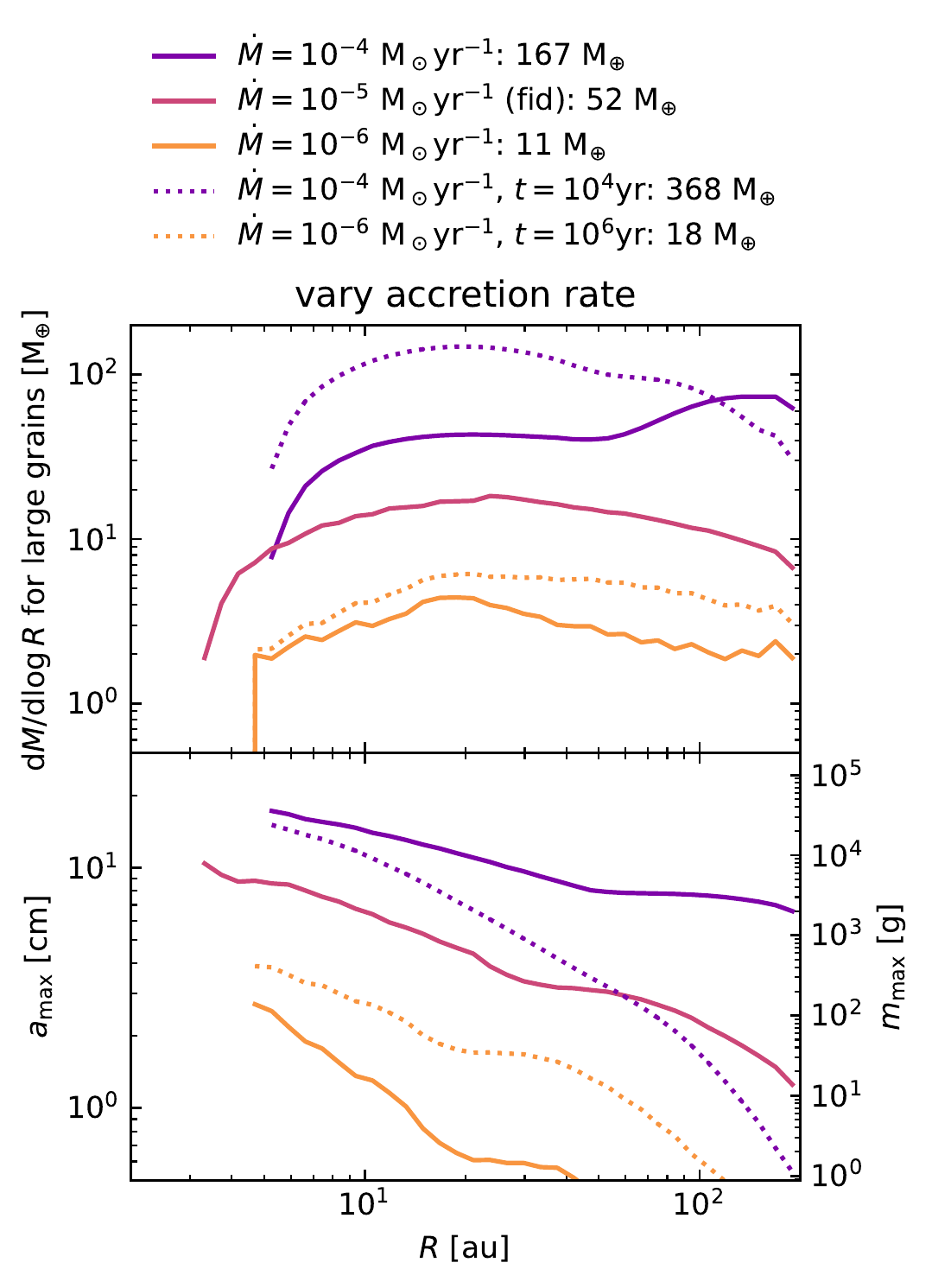}
    \caption{Same as Fig. \ref{fig:vary_Rd}, but now we vary the accretion rate $\dot M$. Solid lines are taken at $t=10^5~{\rm yr}$, and dashed lines are taken at $t = 1~{\rm M}_\odot/\dot M$. The result shows a large scatter in both grain mass and grain size, suggesting that our fiducial model, which does not capture the time evolution of accretion rate during Class 0/I, is best interpreted as an order-magnitude estimate.}
    \label{fig:vary_Mdot}
\end{figure}

In Figs. \ref{fig:vary_Rd} and \ref{fig:vary_Mdot} we run a few different models with different $R_{\rm d}$ and $\dot M$. The results are relatively insensitive to $R_{\rm d}$, mainly because the evolution at large radii is a combination of seeding by outward diffusion and local growth, and is relatively insensitive to the evolution further out.

Meanwhile, the result is more sensitive to accretion rate, with faster accretion producing more mass and higher $m_{\rm max}$ in the sweep-up population.
In particular, for $\dot  M=10^{-4}~{\rm M}_\odot/{\rm yr}$ and $t = 10^4~{\rm yr}$, which roughly represents a fast-accreting Class 0 disk, the mass in large grains reaches several hundred ${\rm M}_\oplus$ (purple dotted line in Fig. \ref{fig:vary_Mdot}).
\revise{We also comment that decreasing the accretion rate causes a large scatter in $m_{\rm max}$ (orange lines in Fig. \ref{fig:vary_Mdot}). This is mainly because low $\dot M$ causes slow sweep-up growth, and the outer disk remains in the linear growth regime where $m_{\rm max}$ depends exponentially on the growth rate and is therefore very sensitive to radius and $\dot M$.}

In reality, we expect the accretion rate to vary from $\sim 10^{-4}~{\rm M}_\odot/{\rm yr}$ in Class 0 to $\sim 10^{-5}~{\rm M}_\odot/{\rm yr}$ in Class I and eventually down to $\sim 10^{-8}~{\rm M}_\odot/{\rm yr}$ during the transition from Class I to Class II.
Therefore, our fiducial model, which only considers a fixed accretion rate, is best interpreted as a rough, order-magnitude estimate.

\subsection{The effect of a more detailed dust coagulation model}\label{sec:results:sensitivity_dust}

\begin{figure}
    \centering
    \includegraphics[scale=0.66]{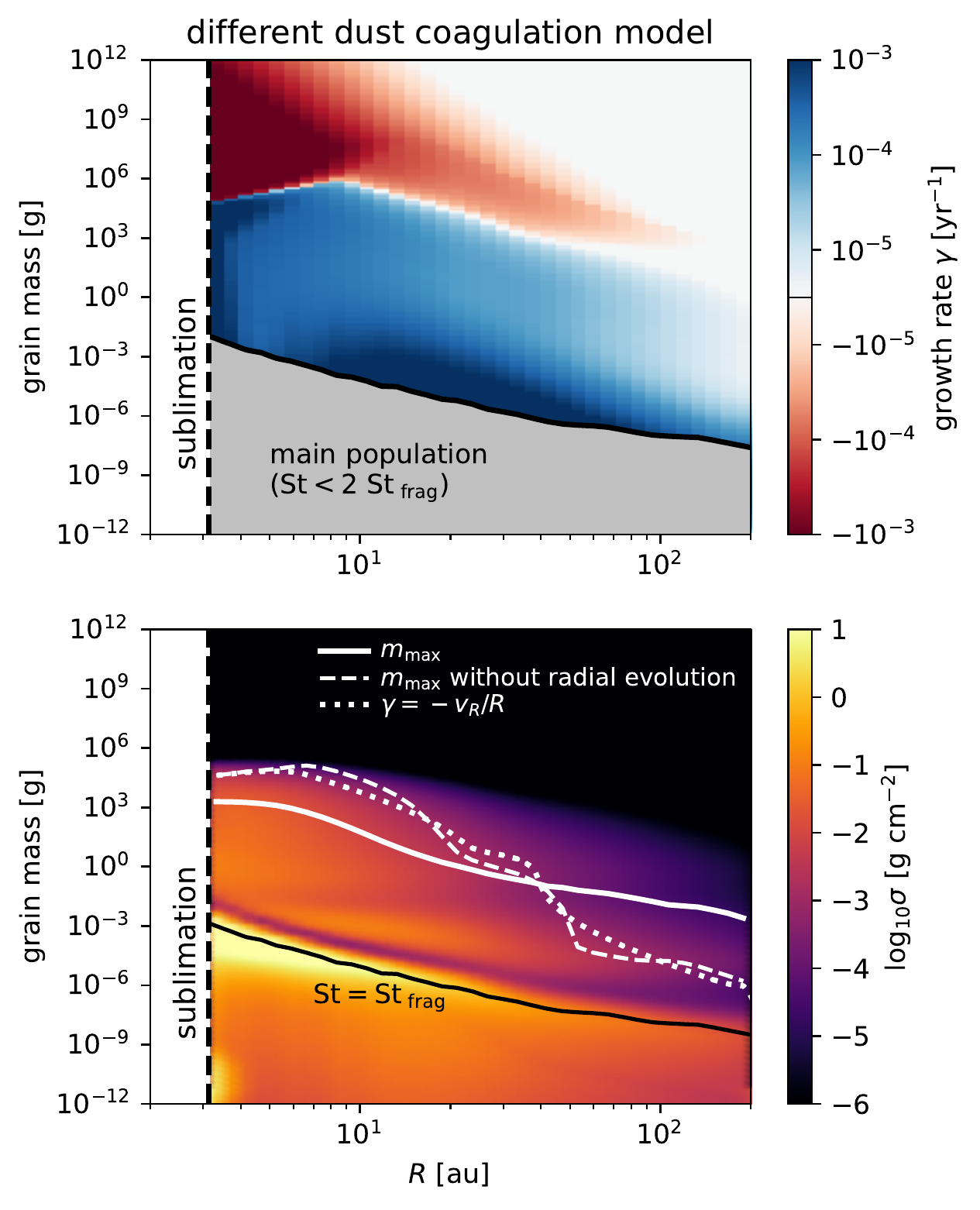}
    \caption{Results for a different grain coagulation model, based on \citet{W12a} and \citet{Kothe2013}. Top panel: sweep-up growth rate, similar to the top panel of Fig. \ref{fig:growth_rate}. Bottom panel: dust mass distribution and $m_{\rm max}$, similar to the bottom panel of Fig. \ref{fig:rad_analysis}. The outcome of sweep-up remains qualitatively similar.}
    \label{fig:W12_K13_coag}
\end{figure}

\begin{figure}
    \centering
    \includegraphics[scale=0.66]{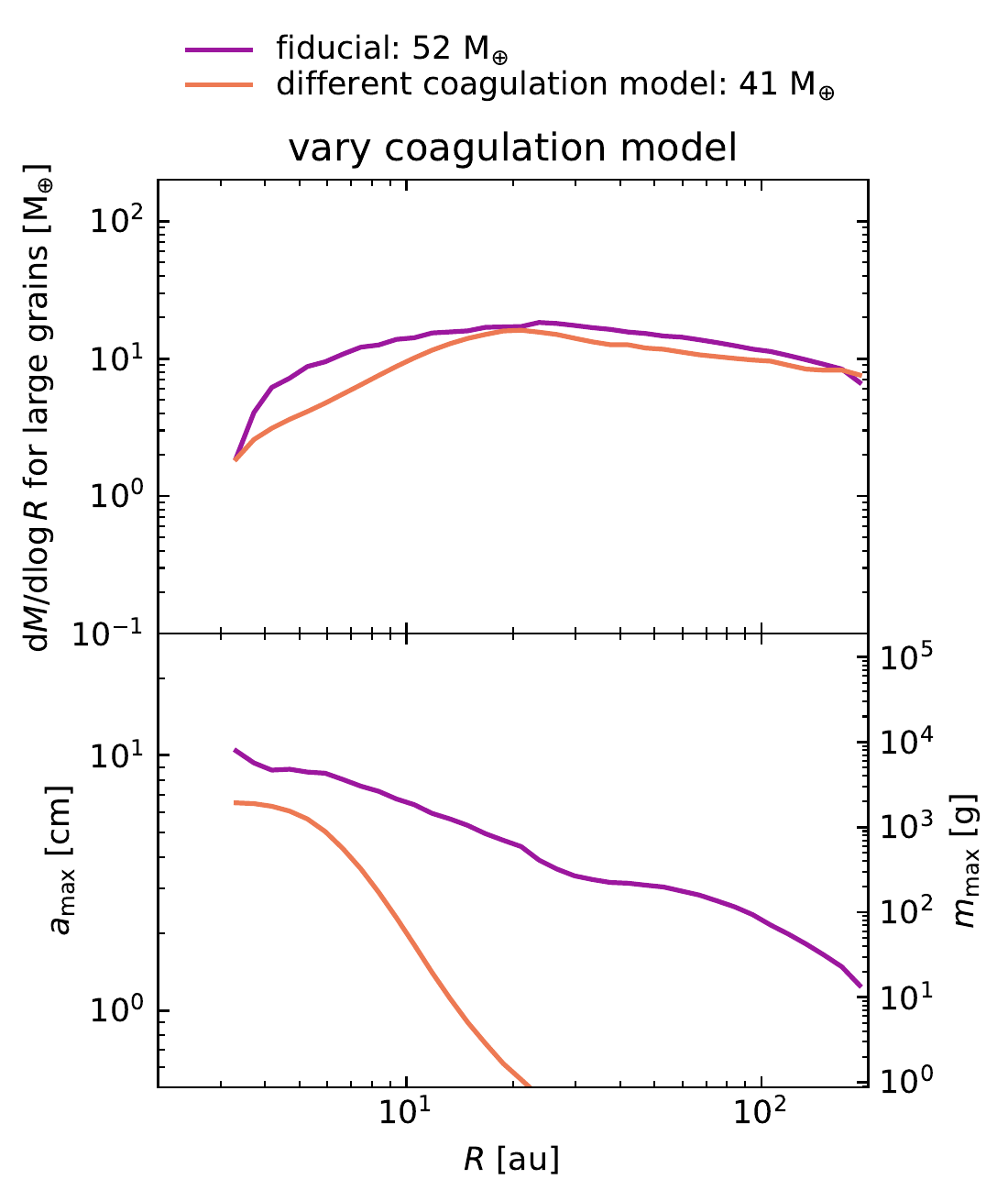}
    \caption{Same as Fig. \ref{fig:vary_Rd}, but compares our fiducial model with a different grain coagulation model, based on \citet{W12a} and \citet{Kothe2013}. While the behaviors are qualitatively similar (cf. Fig. \ref{fig:W12_K13_coag}), The profile of maximum grain size is sensitive to the details of the grain coagulation model.}
    \label{fig:vary_coag}
\end{figure}

Given uncertainties in the details of dust coagulation (cf. Appendix \ref{a:uncertainties}), our fiducial model only includes the minimal complexity required to capture the effect of mass transfer and sweep-up.
Now we estimate the level of uncertainties related to assumptions of the grain coagulation model by comparing our fiducial model with a dust coagulation model based on \citet{W12a} and \citet{Kothe2013}, which includes a mass-dependent fragmentation velocity and a more detailed treatment for the transition between sticking, bouncing, and fragmentation. In particular, this model includes a bouncing regime between sticking and fragmentation that is absent in our fiducial model.

This new model of dust coagulation is based on \citet{W12a}, but with the following differences. First, we include a full treatment of velocity distribution, where we compute collision outcome on a large $v_{\rm rel}$ grid and perform a weighted average assuming a Maxwellian velocity distribution, similar to \citet{W12b}.
Second, we update the treatment on the boundary between sticking and bouncing regimes following a more recent experiment by \citet{Kothe2013}. (The sticking/bouncing model in \citealt{W12a} was based on an earlier work from the same series of experiment.)
Third, in the fragmentation regime of \citet{W12a}, they have variable fragmentation efficiency; here we just follow the \texttt{dustpy} default model and have erosion with $m_{\rm er}=m_{\rm p}$ for large mass ratio and a full fragmentation at lower mass ratio.
Finally, while \citet{W12a} has a more elaborate model of fragment distribution, we stick to the simpler \texttt{dustpy} model discussed in Section \ref{sec:model:dust}.
The last two differences are purely because of technical limitations; following the treatments of \citet{W12a} exactly would require in-depth modification of the \texttt{dustpy} code. We expect the results to be relatively insensitive to these two modifications, because fragmentation/erosion mainly affects the evolution by stopping grain growth, and this effect is relatively insensitive to exactly how much mass is lost per fragmentation (as long as the fractional mass loss is order-unity) or how fragments are distributed.

The result of this new model is summarized in Figs. \ref{fig:W12_K13_coag} and \ref{fig:vary_coag}.
Because of the mass-dependence of the mass transfer rate (cf. \citealt{W12a} Eqs. 8 and 14), the erosion zone is much bigger and the sweep-up rate is no longer approximately constant at given radius. Still, the qualitative evolution of the sweep-up population remains similar to our fiducial model, with the inner disk having sweep-up growth capped by radial motion and the outer disk having large grains deposited by radial diffusion. The amount of mass in the sweep-up population is similar to our fiducial model, but with smaller grain size at larger radii; this is mainly because sweep-up growth is relatively inefficient (compared to radial motion) at large radii and the large grains diffused from smaller radii cannot further grow.

Overall, we find the qualitative behavior of sweep-up to be robust against variations in the grain coagulation model. However, the quantitative details of the outcome (the $a_{\rm max}$ profile, in this example) could be sensitive to the details of the grain coagulation model.

\section{Discussion}\label{sec:discussion}
\subsection{Comparison with sweep-up in Class II disks}\label{sec:discussion:ClassII}

Sweep-up in Class II protoplanetary disks has been studied using very similar dust coagulation models in \citet{W12a, W12b}. They found that although sweep-up can increase grain size by many orders of magnitude, it is challenging to obtain a non-trivial amount of mass (say, $\gtrsim {\rm M}_\oplus$) in the sweep-up population. This is mainly due to two issues: First, sufficiently high growth rate (compared to disk lifetime or viscous timescale) occurs only at very small radii (few au or less) for a typical Class II disk; second, it is difficult to self-consistently generate enough seeds for sweep-up, so the mass in the sweep-up population remains many orders of magnitude below the main population.

In this work, we demonstrate that these issues are naturally resolved by considering younger, Class 0/I protostellar disks.
The high accretion rate of the disk leads to higher temperature and larger effective $\alpha$, which allows sweep-up to operate for a much wider range of disk radii [cf. Eq. \eqref{eq:gamma_simple}].
Furthermore, the turbulent radial diffusion of dust grains helps to seed sweep-up growth at larger radii with large grains formed at smaller radii, resolving the seeding problem there.
Finally, the large mass of gravitationally self-regulated Class 0/I disks (cf. \citealt{Xu22}) ensures that even though the sweep-up population is only a few percent of the total dust mass (Fig. \ref{fig:overview}), it still contains enough mass to be (possibly) relevant for planet formation (Fig. \ref{fig:planet_mass_comparison}).


\subsection{Implications for planet formation}\label{sec:discussion:planet}

\begin{figure}
    \centering
    \includegraphics[scale=0.66]{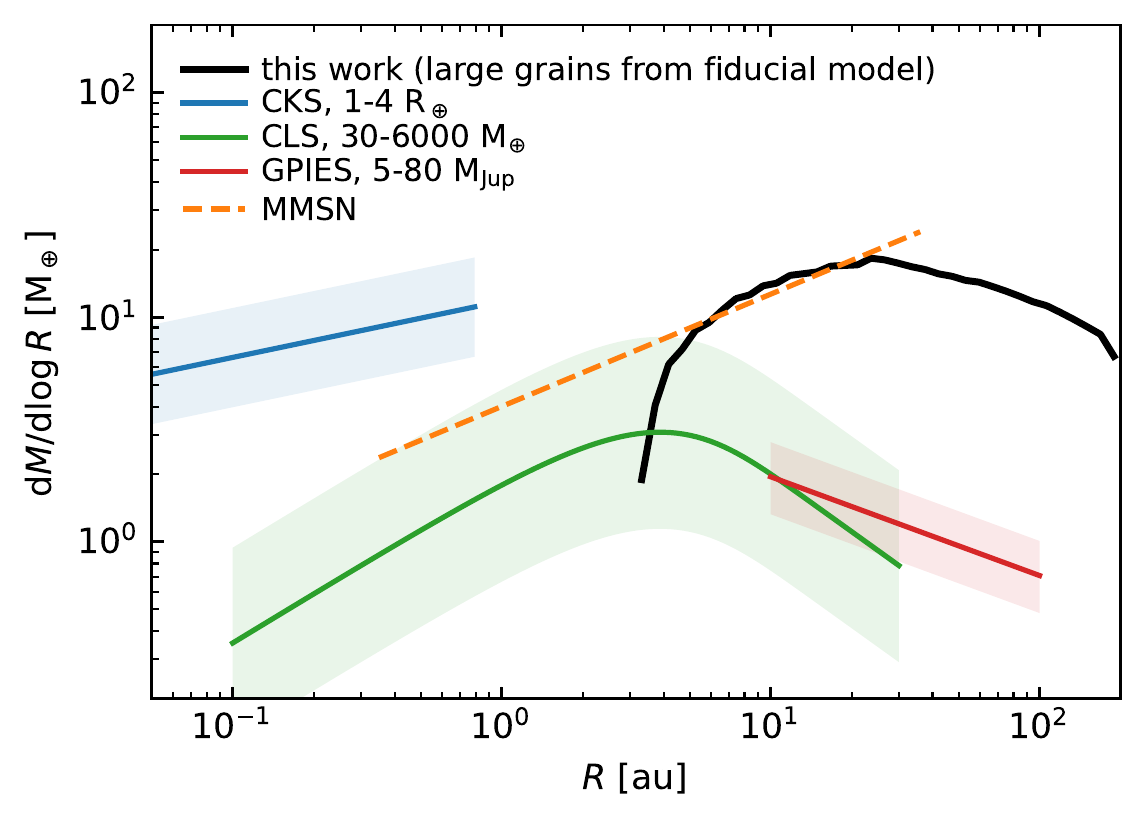}
    \caption{Comparison between the sweep-up population from our fiducial model (black line) and the solid mass budget of planets. Colored lines and shades show observational estimates for different ranges of planet mass (or size) and semi-major axis from three representative observational surveys, CKS (transit, \citealt{Dai2020}), CLS (radial velocity, \citealt{Fulton2021}), and GPIES (direct imaging, \citealt{Nielsen2019}). For CLS and GPIES, which mainly cover giant planets and brown dwarfs, we assume (somewhat arbitrarily) a constant core mass of $30$ and $60~{\rm M}_\oplus$ per planet, respectively. We also show the minimal-mass solar nebula \citep[MMSN,][orange dashed line]{Hayashi1981} for reference.
    This rough comparison demonstrates that in terms of mass budget, it is possible to form all planets from the sweep-up population.
    Testing this possibility, however, would require knowledge of the (highly uncertain) modulations due to disk evolution, planet formation, and planet migration.}
    \label{fig:planet_mass_comparison}
\end{figure}

So far, we have demonstrated that in addition to the main dust population below the fragmentation barrier, protostellar disks host a sweep-up dust population which is much larger in grain size and mass but lower in total surface density. Below we argue that planets may be preferentially fromed from this sweep-up population (compared to the main population) and that it is even possible for all planet solid mass to come from the sweep-up population, with the caveat that future studies are required to access these possibilities more rigorously.

The production of planetesimals (which then form planets) from dust grains can be divided conceptually into three steps \citep{Birnstiel2016}.
First, collisional growth sets some initial grain size. Then, some mechanism of dust concentration or instability leads to an increase in local dust density and dust-to-gas ratio to form dust clumps. Finally, dust clumps collapse under their own gravity into planetesimals.
Here we are interested in how the second step (dust clump formation) is affected by the outcome of collisional growth.
Previous studies have identified a wide variety of candidate mechanisms for dust clump formation.
\revise{In particular, streaming instability \citep{YoudinGoodman2005,Johansen2007} is considered promising by many, and there is a growing literature that characterizes the threshold of instability \citep[e.g.,][]{Carrera2015,LiYoudin2021} and studies the outcome of planetesimal formation \citep[e.g.,][]{Simon2016,Carrera2017,Abod2019,Schafer2022}.
However, given that streaming instability requires a relatively high local dust-to-gas ratio \citep{Carrera2015,LiYoudin2021}, some other mechanism is required to increase the local dust-to-gas ratio to the instability threshold.}
Here this ``other mechanism'' can be one or several of vertical settling, collection by substructure (e.g., pressure bump or vortex), snowline, and (stochastic) turbulent concentration (see reviews in \citealt{Birnstiel2016} and \citealt{Klahr2018}; also see \citealt{Baehr2022})

Given uncertainties in dust dynamics, disk dynamics, and disk evolution, it is often difficult to quantitatively assess the effect of these mechanisms for a given initial dust population produced by collisional growth.
Still, we know qualitatively that for St~$\lesssim 1$, all of the aforementioned mechanisms are benefited by increasing St because it allows dust grains to be less coupled with the gas, promoting dust-to-gas ratio variation and instability.
This makes the sweep-up population (${\rm St}\sim 0.1$--1 in our model) possibly a better candidate for planetesimal formation than the main population (${\rm St}\sim 10^{-4}$--$10^{-2}$ in our model).

Additionally, when we consider planet formation in a more global setting, the $\sim$~kg pebbles formed by sweep-up at large radii ($\gtrsim$ several 10~au) which eventually drifts inward may be a good candidate of pebble flux in young protoplanetary (Class II) disks required by some planet formation models \citep[e.g.,][]{Drazkowska2017}. The exact level of pebble flux produced by the sweep-up population depends on the details of long-term disk evolution, which we hope to address in future studies.

On the other hand, the low surface density of the sweep-up population tends to make planetsimal formation more difficult, because it requires a larger factor of dust density enhancement before reaching the streaming instability threshold \citep{LiYoudin2021}. This is the main reason why in our model (which only considers Class 0/I and does not consider any mechanism of dust concentration) we do not expect streaming instability to occur for the sweep-up population, even though its St is in the most preferable range. Whether the disadvantage of lower surface density will overturn the advantage of higher St is still an open question.

We also comment that the total mass of the sweep-up population is comparable to or larger than the observational estimates of the total solid mass of (solar or extrasolar) planets, as demonstrated by the comparison in Fig. \ref{fig:planet_mass_comparison}.
This implies that in the limit where planetesimal formation is much more efficient for the sweep-up population than the main population, all planets may be formed from the sweep-up population.

In summary, our discussion above suggests that the sweep-up population is \textit{possibly} an important piece in the theory of planet (or planetesimal) formation, because it is probably easier to form planetesimals from the sweep-up population (which has larger St) than from the main population and the sweep-up population contains enough mass to form all planets.
Meanwhile, assessing the role of the sweep-up population on planet formation in a more quantitative and reliable manner requires many further studies, especially on how mechanisms of dust clump formation work on this sweep-up population, and whether the advantage of larger grain size can overcome the disadvantage of lower dust surface density (compared to the main population).


\section{Conclusion}\label{sec:conclusion}
In this paper we investigate the outcome of collisional dust growth in Class 0/I protostellar disks and discuss its implication on planet formation.
In a Class 0/I disk, while most grains cannot grow past the fragmentation barrier, a small population of grains can grow to much larger pebbles via the mass transfer from much smaller grains during high-velocity collisions (sweep-up).
In our fiducial simulation, this sweep-up dust population contains $\revise{52}~{\rm M}_\oplus$ of mass in large pebbles well above the fragmentation barrier with maximum mass reaching $\sim$kg and above (Fig. \ref{fig:overview}).
We discuss the linear growth and saturation of sweep-up, and find that the growth rate scales with disk temperature and effective viscosity (Section \ref{sec:results:growth}), and the saturation of sweep-up is due to finite system lifetime, fragmentation with similar-sized grains in the sweep-up population, or inward radial motion (advection and drift), depending on the location in the disk (Section \ref{sec:results:norad} and \ref{sec:results:rad}). In particular, the sweep-up population spans a wide range of disk radii because turbulent diffusion transport large grains in the inner disk to the outer disk and seeds further growth there.
Under reasonable variation of disk parameters and grain coagulation model, the sweep-up population emerges robustly, although the total surface density and maximum grain size of the sweep-up population can vary by a factor of a few in both directions (Section \ref{sec:results:sensitivity_disk} and \ref{sec:results:sensitivity_dust}).
\revise{Still, there remain several uncertainties in the dust coagulation model (Appendix \ref{a:uncertainties}) that need to be addressed in future studies.}

We explain why sweep-up is more effective in Class 0/I than in Class II (Section \ref{sec:discussion:ClassII}), and argue that planetesimals may preferentially form from the sweep-up population compared to the main population (Section \ref{sec:discussion:planet}) because their higher (but still $\lesssim 1$) Stokes number make them more available to mechanisms of dust clump formation.
In the most extreme case, it is even possible for the whole planet population to be formed from this sweep-up population of pebbles (Fig. \ref{fig:planet_mass_comparison}).
However, it remains uncertain whether the advantage of higher St can overcome the disadvantage of lower surface density. 

This work is only a first step towards understanding the first stages of planet formation in Class 0/I disks, and we hope our results inspire future studies that consider whether known planetesimal formation mechanisms could be more effective in Class 0/I disks than in Class II disks.
It would also be important to access systematically the role of gravitational instability (which is expected for typical Class 0/I disks) in grain growth and planetesimal formation.

WX thanks Eugene Chiang, Hubert Klahr, Jiayin Dong, Rixin Li, and Sebastian Stammler for helpful discussions. PJA acknowledges support from NASA TCAN award 80NSSC19K0639.

\software{
Dustpy \citep{dustpy},
Matplotlib \citep{matplotlib}
}

\bibliography{XA22}{}
\bibliographystyle{aasjournal}
\appendix
\section{Uncertainties in modeling collision outcome}\label{a:uncertainties}
In this appendix we briefly review two major uncertainties in modeling the outcome of grain-grain collision, and discuss how they affect the results of this paper and other studies in the literature.

\subsection{Fragmentation threshold}\label{a:uncertainties:frag}
The fragmentation threshold has long been a major uncertainty in collisional grain growth.
Lab experiments and numerical simulations of grain-grain collision with different grain mass, grain material, and compactness (or porosity) find a wide range of fragmentation threshold velocities ranging between 0.2~m/s to several 10~m/s \citep[e.g.,][]{BlumWurm2008,Wada2009,Beitz2011}. Early studies of grain growth generally assume $v_{\rm frag} = 1$~m/s \citep[e.g.,][]{B10}, following lab experiments of dust aggregates. More recently, assuming a higher fragmentation threshold of $v_{\rm frag} = 10$~m/s based on simulations of icy grains is becoming more common \citep[e.g.,][]{Stammler2019}. It is also worth noting that directly growing grains into pebbles that can be affected by drift (which provides a pebble flux for the inner disk) and clumping mechanisms requires both high fragmentation velocity ($\sim 10$~m/s) and relatively low level of turbulence ($\alpha\lesssim 10^{-3}$). The latter is much lower than the level of turbulence in the outer part of our Class 0/I disk model.

On the other hand, the choice of fragmentation threshold is less important for the outcome of sweep-up. Increasing the fragmentation threshold in our model to 10~m/s beyond the snowline would barely affect the sweep-up growth rate, because it is insensitive to the size of the small grains as long as erosion is weak [Eqs. \eqref{eq:eps_net}, \eqref{eq:gamma_simple}; but also see Appendix \ref{a:uncertainties:mt}].
A larger fragmentation threshold also affects the yield of sweep-up by providing larger seeds, but this effect would probably be marginal given that the seeding is dominated by the outward diffusion of large grains formed at smaller radii (and often within the snowline).

\revise{
\subsection{Erosion and mass transfer efficiency}\label{a:uncertainties:mt}

The efficiency of mass transfer and erosion directly affect the outcome of sweep-up.
Here a major uncertainty lies in how lab experiment results (see a review in \citealt{Blum2018}) should be extrapolated in grain mass, relative velocity, and grain material (including porosity/compactness).
Below we summarize a few trends in lab experiments and numerical simulations and discuss how they could affect the erosion and mass transfer rates in our calculation.

When the projectile is relatively large ($\gtrsim$mm), the compactness of the grains is probably an important driver of mass transfer. At relative velocity $\lesssim$ few 10~m$/$s, experiments suggest that the the outcome of collision is very sensitive to porosity difference $\Delta\phi$.
\citet{Paraskov2007} experiment with several different materials, and find that erosion happens only when the projectile is as compact as or more compact than the target ($\Delta\phi\leq 0$). In particular, they find that equal compactness projectile and target leads to a weak mass loss, but slightly increasing the compactness of the target (by facing the projectile with the slightly more compact bottom side of the target) instead leads to mass gain.
\citet{Beitz2011} study the collision of equal-sized cm aggregates, and find that whenever there is a finite porosity difference, only the more porous aggregate fragments (with some of the fragmented mass sticking onto the other aggregate) and the more compact aggregate remains fully intact (cf. Fig. 7 of \citealt{Beitz2011}).
This happens even when $\Delta\phi$ is at tiny as $\sim 0.025$; and in general the mass transfer efficiency scales as $\Delta \phi^2$ (Eq. 8 of \citealt{Beitz2011}).
At higher relative velocity ($\gtrsim$ few 10~m$/$s), the collision outcome also depends on the relative velocity and projectile size, with faster relative velocity and larger projectile promoting erosion. This is demonstrated by the experiments in \citet{TeiserWurm}, which is used to calibrate our erosion parameterization Eq. \eqref{eq:m_er} (cf. Fig. 4 in \citealt{W12a}).

For small projectiles (1 -- several 10~$\mu$m), the projectile mass dependence reverses and smaller projectiles lead to stronger erosion. Meanwhile, the dependence on relative velocity and compactness remains qualitatively similar, with higher relative velocity promoting erosion and more compact target suppressing erosion. \citet{Schrapler2018} perform high-velocity collision experiments with $\Delta\phi\approx 0$ and find that the boundary between erosion and accretion is given by
\begin{equation}
\left(\frac{r_{\rm p}}{2\times 10^{-5}~{\rm m}}\right)^{-0.62} \frac{v_{\rm rel}}{15~{\rm m/s}} \sim 1.
\end{equation}
Simulations in \citet{Seizinger2013} show the same trend; they also demonstrate that that choosing a more compact target only slightly increases the threshold velocity for erosion. In other words, target compactness seems to play a less significant role in this regime.

These trends could impact our results in two ways.
First, we have likely underestimated the erosion efficiency for small grains, because Eq. \eqref{eq:m_er} does not capture the increase in erosion efficiency for very small grains. Comparing Eq. \eqref{eq:m_er} against the monomer erosion rate in \citet{Schrapler2011}, we find that Eq. \eqref{eq:m_er} could underestimate erosion by an order of magnitude for the smallest grains.
This might suppress sweep-up growth in the outer part of the disk, where the typical grain size in the main population is small and turbulent velocity is large.

On the other hand, we have probably overestimated erosion and underestimated mass transfer for larger projectiles. Our mass transfer efficiency parametrization Eq. \eqref{eq:m_mt} assumes $\Delta\phi=0.1$ and our erosion efficiency parametrization Eq. \eqref{eq:m_er} is based on experiments with $\Delta\phi=0$. In reality, the projectiles from the main population are mainly grown from low-velocity sticking and are expected to be relatively porous; meanwhile, the targets in the sweep-up population grow by many high-velocity collisions and are probably more compact. As a result, it might be reasonable to ignore erosion altogether for larger projectiles, which would eliminate the erosion zone in Fig. \ref{fig:growth_rate} and Fig. \ref{fig:W12_K13_coag} and allow grains at small radii to easily grow to ${\rm St} \sim 1$. The mass transfer efficiency may also be larger if the difference in porosity turns out to be larger than our assumed value of $0.1$.
}

\end{document}